# When Cultures Meet: Modelling Cross-Cultural Knowledge Spaces

Anneli HEIMBÜRGER
*University of Jyväskylä*
*Faculty of Information Technology*
*Information Technology Research Institute*
*P.O. Box 35 (Agora)*
*FIN-40014 University of Jyväskylä, Finland*
anneli.heimburger@titu.jyu.fi

**Abstract.** Cross-cultural research projects are becoming a norm in our global world. More and more projects are being executed using teams from eastern and western cultures. Cultural competence might help project managers to achieve project goals and avoid potential risks in cross-cultural project environments and would also support them to promote creativity and motivation through flexible leadership. In our paper we introduce an idea for constructing an information system, a cross-cultural knowledge space, which could support cross-cultural communication, collaborative learning experiences and time-based project management functions. The case cultures in our project are Finnish and Japanese. The system can be used both in virtual and in physical spaces for example to clarify cultural business etiquette. The core of our system design will be based on cross-cultural ontology, and the system implementation on XML technologies. Our approach is a practical, step-by-step example of constructive research. In our paper we shortly describe Hofstede's dimensions for assessing cultures as one example of a larger framework for our study. We also discuss the concept of time in cultural context

## 1. Introduction

The Internet and ubiquitous technology have opened up new possibilities for us to promote research and development projects as well as our business activities to new geographical locations and cultures. It is almost as easy to work with people remotely as it is to work face-to-face. Cross-cultural communication is more and more the new norm for our collaborative operations.

Increasingly, businessmen, project managers, researchers and other professionals are becoming involved in international negotiations and meetings. The meetings can for example be international business meetings or international research project meetings. In addition to meeting agenda, participants also share culturally integrated space. Sometimes it can be difficult to understand culture dependent behavior of other parties during a meeting. By understanding some of the main cultural dimensions and by adjusting to cultural differences, people can face the challenge and become better negotiators and project managers on behalf of their companies and research organizations.

The objective of our research project is to design and implement an information system – a cross-cultural knowledge space – that provides cultural assistant for people attending in cross-cultural meetings or for people working in cross-cultural projects [8]. The system can be used personally or collaboratively both in virtual spaces and in physical spaces.

The contribution of the paper is to:
- introduce a cultural ontology based approach to construct an information system that could promote communication and mutual understanding in cross-cultural collaborative research project environments, especially between eastern and western cultures
- describe Hofstede's framework for cultural dimensions which is based on questionnaire study in 74 countries and on statistical analysis of the survey data
- discuss the concept of time in cultural context as an essential issue of time-based project management functions.

The term "culture" is used in our paper as it is defined in [11]: "Culture is a collective phenomenon, because it is shared with people who live or lived within the same social environment, which is where it was learned. Culture consists of the unwritten rules of the social game. It is the collective programming of the mind that distinguishes the member of one group or category of people from others". The concept "cross-cultural" is used in the paper to describe comparative knowledge and studies of a limited number of cultures. For example, when examining negotiation manners or attitudes towards time in Finland and in Japan than that is a cross-cultural study. The concept "knowledge space" in cross-cultural context is used to describe personal and collaborative information systems both in virtual worlds on the fixed or ubiquitous Web and in physical worlds like in meeting rooms.

The remainder of the paper is organized as follows. In Section 2, we describe a framework for assessing cultures with five cultural dimensions. In Section 3 we discuss the concept of time in cultural context. In Section 4, we introduce an idea for constructing an information system that supports cross-cultural communication in virtual and/or in physical space. The system is based on cultural ontology. We also present technological tools for and their roles in the implementation. Section 5 is reserved for conclusions and issues for further steps.

## 2. A Framework for Cultural Dimensions

All of us, who are working, for example in international research projects, are involved – in addition to the subject of the project itself – in another kind of development process. Cultural competence [15] is a developmental process that evolves step-by-step over an extended period. Both individuals and organizations are at various levels of awareness, knowledge and skills on the cultural competence continuum. Cultural competence is about respecting cultural differences and similarities.

There exist several studies for assessing cultures [11, 15]. These studies consider relations between people, motivational orientation, orientation towards risks, definition of self and others, attitudes to time, and attitudes to environments. Hofstede's framework for assessing cultures is one of the widely used frameworks [10, 11]. Hofstede's approach proposes a set of cultural dimensions along which dominant value systems can be ordered. These value systems affect human thinking, feeling, and acting, and the behavior of organizations and institutions in predictable ways. The framework consists of five dimensions: individualism/collectivism, power distance, masculinity/femininity, uncertainty avoidance and long-term orientation/short-term orientation (Table 1). All dimensions are generalizations and individuals may vary from their society's descriptors.

Hofstede's metrics provides on interesting, larger framework for our study. In addition to this larger framework there are several culture dependent characteristics which persons can face in their everyday working life. One example is communication style which can be indirect, paraverbal and/or nonverbal [18]. Nor should the role of business domain and organization specific cultures be underestimated. Awareness of cultural dimensions together with culture-specific characteristics could help people to develop their cultural competence.

**Table 1.** Summary of cultural dimensions according to Hofstede's study

| Dimension | Description of the dimension |
| --- | --- |
| Individualism/ Collectivism | Individualism/Collectivism describes the extent to which a society emphasizes the individual or the group. Individualistic societies encourage their members to be independent and look out for themselves. Collectivistic societies emphasize the group's responsibility for each individual. |
| Power distance | Power distance describes the extent to which a society accepts that power is distributed unequally. When power distance is high, individuals prefer little consultation between superiors and subordinates. When power distance is low, individuals prefer consultative styles of leadership. |
| Masculinity/Femininity | Masculinity/Femininity refers to the values more likely to be held in a society. Masculine societies are characterized by an emphasis on money and things. Feminine cultures are characterized by concerns for relationships, nurturing, and quality of life. |
| Uncertainty avoidance | Uncertainty avoidance refers to the extent that individuals in a culture are comfortable (or uncomfortable) with unstructured situations. Societies with high uncertainty avoidance prefer stability, structure, and precise managerial direction. In low uncertainty avoidance societies are comfortable with ambiguity, unstructured situations, and broad managerial guidance. |
| Long-term/short-term orientation | Long-term/short-term orientation refers to the extent to which a culture programs its members to accept delayed gratification of their material, social, and emotional needs. Business people in long-term oriented cultures are accustomed to working toward building strong positions in their markets and do not expect immediate results. In short-term oriented cultures the "bottom line" (the results of the past month, quarter, or year) is a major concern. Control systems are focused on it and managers are constantly judged by it. |

The scores of cultural dimensions in different countries according to Hofstede's research are given in [12]. The survey is extensively described in [10]. The figures should not be taken literally. However they provide interesting information because they show differences in answers between groups of respondents.

### 3. Time in Cultural Context

Time is seen in a different way by eastern and western cultures and even within these groupings temporal culture differs from country to country. Also temporal identities of different organizations and teams in organizations may vary. In cultural context, there exist two general time models: linear and cyclic [15]. In linear time model (Figure 1a) past time is over, present time can be seized and parceled and make it work for the immediate future. One task is carried out at time. For example, Scandinavian people are essentially linear-active, time-dominated and monochronic. They prefer to do one thing at a time, concentrate on it and do it within a scheduled timetable. Southern Europeans are more multi-active and polychronic. Monochronic cultures differ from polychronic cultures in that the former encourage a highly structured, time-ordered approach to life and the latter a more flexible, indirect approach, based more upon personal relationships than scheduled commitments.

In many Asian countries time has traditionally been considered as cyclic. For example, the Japanese traditional temporal culture can be presented by the Makimono model of time (Figure 1b) [7]. In Makimono time, the future flows into the present, just as the past does. The present is a period that links the region of the past with the world of the future. Nowadays linear time model has also been integrated into Japanese society.

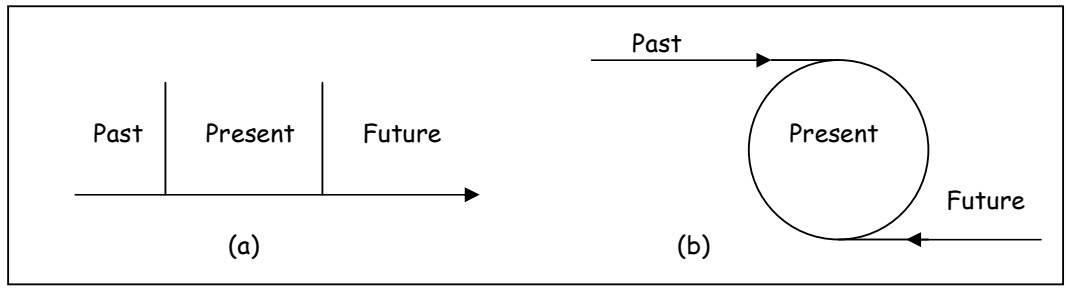

**Figure 1**. Linear time model and cyclic time model according to Makimono time pattern. Makimono takes its name from the makimono, a picture story or writing mounted on paper and usually rolled into a scroll.

Cross-cultural projects involve teams and individuals with different concepts of time, and therefore a completely different frame of mind as far as planning, scheduling, punctuality and project deadlines are concerned. Tensions may arise quite easily. In such a case, it is the task of the project manager, on the basis of his/her cultural competence, to make sure such different attitudes do not become the source of major misunderstandings. Time contexts in project management are discussed more detailed in [9].

## 4. Towards a Cross-Cultural Ontology

The development process of cultural competence of project managers and project teams could be supported by culture-sensitive information systems both in virtual and in physical environments. In our system we first construct a cross-cultural ontology which will be the basis of the system.

An ontology is the result of an attempt to formulate an exhaustive and rigorous conceptual schema about a certain domain. The domain does not have to be the complete knowledge of that topic, but an interesting part of it decided by the creator of the ontology. In our approach, the cultural dimensions discussed in the Section 2 can be grouped into three categories: relations between people, motivational orientation and attitudes towards time. These categories can be complemented with an application category which includes cross-cultural applications such as project negotiations [2, 16] and time-based project management. The four categories form the first hierarchy level of a cross-cultural ontology (Figure 2):

(individualism, collectivism) ∈ **Relations between people**
(masculinity, femininity, uncertainty avoidance, power distance) ∈ **Motivational orientation**

(long-term orientation, short-term orientation, linear time, cyclic time) ∈ **Attitudes towards time**
(project negotiations, project management) ∈ **Applications**

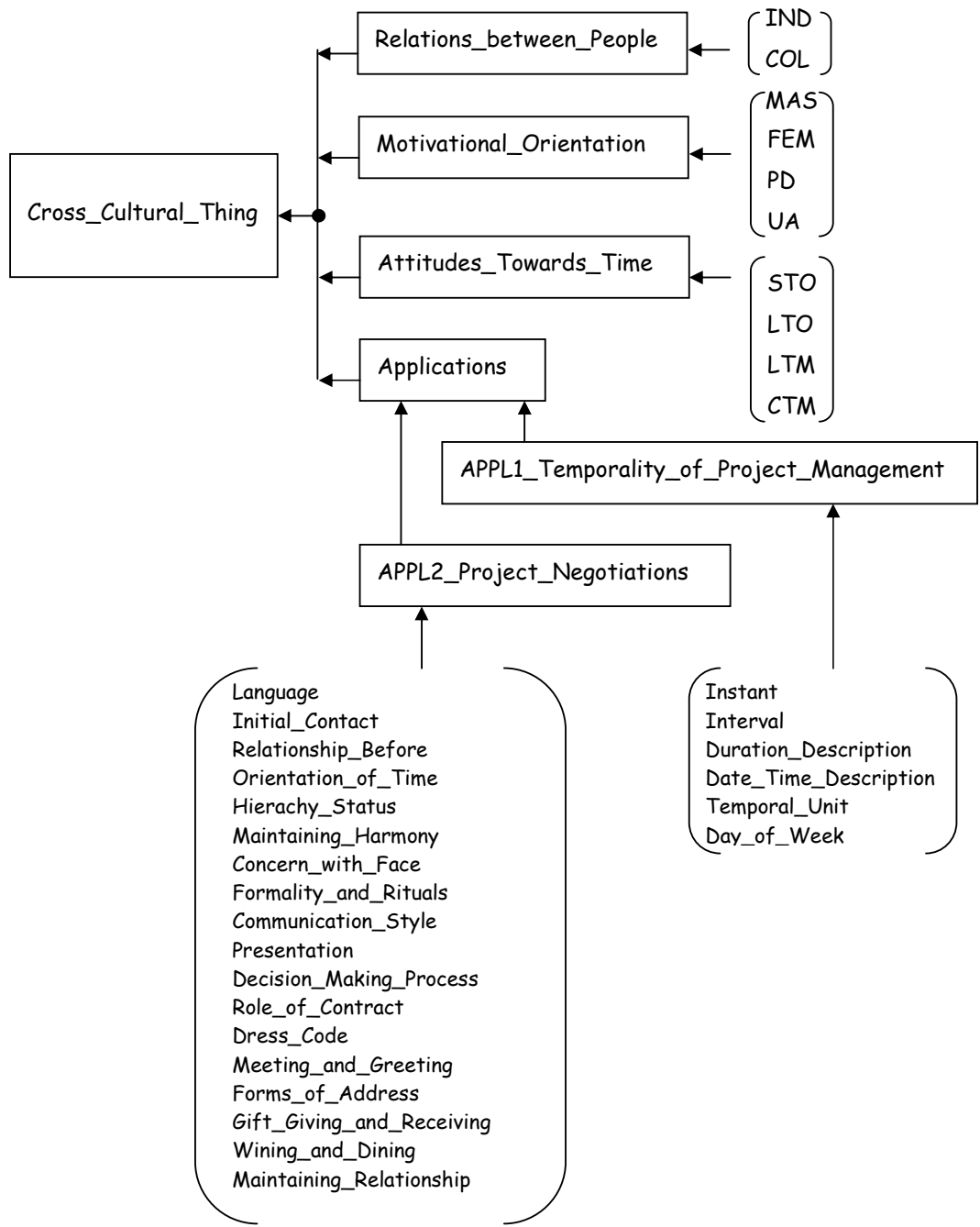

**Figure 2**. Cross-cultural ontology can be associated to cultural knowledge that is represented in XML documents. The case cultures in our project are Finnish and Japanese. For example in application concerning project negotiations there can be a collection of XML documents describing a Japanese Negotiator and a Finnish Negotiator. The following abbreviations are used in the figure: Individualism = IND, Collectivism = COL, Masculinity = MAS, Feminity = FEM, Power Distance = PD, Uncertainty Avoidance = UA, Long-Term Orientation = LTO, Short-Term Orientation = STO, LTM = Linear Time Model and CTM = Cyclic Time Model.

The idea of the system design is that it can be used both in virtual and in physical environments i.e. (a) as a personal assistant via mobile devices, (b) as a collaborative assistant in meeting rooms and (c) as a personal/collaborative assistant in a virtual project space. Basically, the same idea can be applied for example to cross-cultural business meetings, education, tourism, medical and social services. The functions of essential technologies for implementation are shortly summarized in Table 2.

**Table 2.** Essential technologies for constructing cross-cultural knowledge spaces

| Technology | Functions |
| --- | --- |
| Ubiquitous and Context Aware Computing | Ubiquitous computing refers to a new computing paradigm that focuses on offering user-friendly information services - anywhere and anytime [20]. The core function is to support users by means of a cross-cultural knowledge space that is aware of their presence and cultural context. Context is any information that can be used to characterize the situation of an entity. An entity can be a person, a place, a space, time or an object that is considered relevant to the interaction between a user and an application. The system is context-ware if it uses context to provide relevant information and/or services to the user where relevancy depends on the user's task or situation [3, 4]. Examples of contexts in cross-cultural environments are nationality (static situation), location and time (dynamic situation), preferences (static intension) and joint project activities (dynamic intension). |
| Web Ontology Language (OWL) | OWL is a markup language for publishing and sharing data using ontologies on the Internet [24]. OWL is used to formulate a conceptual schema for cultural entities. |
| OWL-Time | OWL-Time presents an ontology of temporal concepts [23]. The ontology provides a vocabulary for expressing facts about topological relations among instants and intervals, together with information about durations, date and time. OWL-Time is used as a basis time ontology in cross-cultural time-based project management applications. |
| N-ary relations | In Semantic Web languages, such as RDF and OWL, a property is a binary relation. It is used to link two individuals or an individual and a value. However, in some cases, the natural and convenient way to represent certain concepts is to use relations to link an individual to more than just one individual or value. These relations are called n-ary relations [22, 25]. In our ontology we need for example to represent multicultural properties of an object. |
| Kansei Information Processing | Kansei is an ability that allows humans to solve problems and process information in a personal way. In every action performed by a human being, traces of his/her Kansei can be noticed, as well as his/her way of thinking and solving problems. Kansei is related both to problem solving tasks and to information analysis and synthesis. [1, 5, 6]. In the design of information systems, the concept of Kansei is related to data definition and data retrieval [13, 14]. In our |

| | research we study how cultural dependent semantic attributes could be added to Kansei information processing and thus how culture-sensitive information retrieval can be supported. |
|---|---|
| XML for Emotions | An important function in cross-cultural virtual spaces is to express emotions. XML based language for emotions could be one approach for expressing emotional functions. |
| XML Topic Maps (XTM) | In topic maps [21, 27], three constructs are provided for describing the subjects represented by the topics: topic names, occurrences, and associations. Topic can be typed. Occurrences relate topics to the information they are relevant to. *Table 2 continues …* |
| XML Topic Maps (XTM) | *Table 2 continues …* Occurrences use URI addresses to identify the information resources, such as XML documents, being connected to the topic. Associations represent relationships between topic, and like occurrences they can be typed. The relationships in traditional classification schemes have little semantic content, whereas in topic maps one generally tries to make the typing of associations as specific as possible. In our project, Topic Map approach is used to design the user interface. |
| Virtual spaces | An intelligent virtual space platform will be selected for the project |
| Radio Frequency Identification | RFID is an automatic identification method, relying on storing and remotely retrieving data using devices called RFID tags or transponders [17, 19]. A typical RFID solution consists of a data gatherer, RFID reader, and a data carrier (RFID tag) that is attached to an item (mobile device) or location (meeting room). Meeting rooms in organizations can be regarded as being context-sensitive areas and appropriately equipped by means of RFID technology for cross-cultural applications. |

## 5. Conclusions and Next Steps

In our paper we introduced an idea for constructing an information system that could support cross-cultural communication and project management functions in collaborative virtual or physical spaces. Our system will be designed by means of a cross-cultural ontology and will be based on XML and agent [26] technologies. The plan for our next steps is:
      Phase 1: System design, demonstrator implementation, testing in laboratory
      Phase 2: Qualitative evaluation in selected test sites
      Phase 3: Focus our design towards cross-cultural agent (CCA) applications.
    Cultural competence can be regarded as a set of congruent functions such as behaviors, attitudes, and policies that work in an information system and/or among professionals and enable the system and the professionals to work effectively in cross–cultural situations. From operational point of view, cultural competence is the integration and transformation of knowledge about cultures, groups of people and individuals into specific standards, policies, practices, and attitudes. These are used in appropriate cultural settings to increase the quality and context-sensitivity of information systems.
    Projects that use effective cross-cultural human-computer systems could provide a source of learning experiences and innovative thinking to enhance the competitive position of the participating organizations.

**Acknowledgements**

This is the author's version of the work. The definite version was published in Heimbürger, A. 2008. When Cultures Meet – Modelling Cross-Cultural Knowledge Spaces. In: Jaakkola, H., Kiyoki, Y. and Tokuda, T. (eds.) Frontiers in Artificial Intelligence and Applications, Vol. 166, Information Modelling and Knowledge Bases XIX. Amsterdam: IOS Press. Pp. 314 – 321.

We express our deep thanks to the Satakunta High Technology Foundation and to the Scandinavia-Japan Sasakawa Foundation for funding the preliminary phase of our research project.